\documentclass[a4paper]{jpconf}
\usepackage{graphicx}
\begin{document}
\title{LUMINEU: a search for neutrinoless double beta decay based on ZnMoO$_4$ scintillating bolometers}

%\vskip 0.3cm

% The LUMINEU and the EDELWEISS collaborations

\author{E.~Armengaud~$^{1}$, Q.~Arnaud~$^{2}$, C.~Augier~$^{2}$,
A.~Beno$\mathrm{\hat{i}}$t~$^{2}$,
A.~Beno$\mathrm{\hat{i}}$t~$^{3}$, L.~Berg\'e~$^{4}$,
 R.S.~Boiko~$^{5}$, T.~Bergmann~$^{6}$, J. Bl\"{u}mer~$^{7,8}$, A.~Broniatowski~$^{4}$, V.~Brudanin~$^{9}$, P.~Camus~$^{3}$, A.~Cazes~$^{2}$,
 M.~Chapellier~$^{4}$, F.~Charlieux~$^{2}$, D.M.~Chernyak~$^{5}$, N.~Coron~$^{10}$, P.~Coulter~$^{11}$, F.A.~Danevich~$^{5}$, T. de Boissi\`ere~$^{1}$,
 R.~Decourt~$^{12}$, M.~De~Jesus~$^{4}$, L.~Devoyon~$^{13}$, A.-A.~Drillien~$^{4}$, L.~Dumoulin~$^{4}$, K.~Eitel~$^{8}$, C.~Enss~$^{14}$, D. ~Filosofov~$^{9}$,
 A.~Fleischmann~$^{14}$, N.~Foerster~$^{7}$, N.~Fourches~$^{1}$, J.~Gascon~$^{2}$, L.~Gastaldo~$^{14}$, G.~Gerbier~$^{1}$, A.~Giuliani~$^{4,15,16}$,
 D.~Gray~$^{1}$, M.~Gros~$^{1}$, L.~Hehn~$^{8}$, S.~Henry~$^{11}$, S. Herv\'e~$^{1}$, G.~Heuermann~$^{7}$, V.~Humbert~$^{4}$,
 I.M.~Ivanov~$^{17}$, A.~Juillard~$^{2}$, C.~K\'ef\'elian~$^{2,7}$, M.~Kleifges~$^{6}$, H.~Kluck~$^{7,18}$, V.V.~Kobychev~$^{5}$,
 F.~Koskas~$^{13}$, V.~Kozlov~$^{8}$, H.~Kraus~$^{11}$,
 V.A.~Kudryavtsev~$^{19}$, H.~Le~Sueur~$^{4}$, M. Loidl~$^{20}$, P.~Magnier~$^{1}$, E.P.~Makarov~$^{17}$, M.~Mancuso~$^{4,15}$, P.~de~Marcillac~$^{4}$, S.~Marnieros~$^{4}$,
 C.~Marrache-Kikuchi~$^{4}$, A.~Menshikov~$^{6}$, S.G.~Nasonov~$^{17}$, X.-F.~Navick~$^{1}$, C. Nones~$^{1}$,
 E.~Olivieri~$^{4}$, P.~Pari~$^{21}$, B.~Paul~$^{1}$, Y.~Penichot~$^{1}$, G.~Pessina~$^{16,22}$, M.C.~Piro~$^{4}$, O.~Plantevin~$^{4}$,
 D.V.~Poda~$^{4,5}$, T.~Redon~$^{10}$, M.~Robinson~$^{19}$,  M.~Rodrigues~$^{20}$, S.~Rozov~$^{9}$,
 V.~Sanglard~$^{2}$, B.~Schmidt~$^{8}$, S.~Scorza~$^{7}$ V.N.~Shlegel~$^{17}$, B.~Siebenborn~$^{8}$,
 O.~Strazzer~$^{13}$, D.~Tcherniakhovski~$^{6}$, M.~Tenconi~$^{4}$, L.~Torres~$^{10}$, V.I.~Tretyak~$^{5,23}$, L.~Vagneron~$^{2}$,
 Ya.V.~Vasiliev~$^{17}$, M.~Velazquez~$^{12}$, O.~Viraphong~$^{12}$, R.J.~Walker~$^{8}$, M.~Weber~$^{6}$,
 E.~Yakushev~$^{9}$, X.~Zhang~$^{11}$ and V.N.~Zhdankov~$^{24}$
 }

\address{$^{1}$~CEA, Centre d'Etudes Saclay, IRFU, 91191 Gif-Sur-Yvette Cedex, France

 $^{2}$~IPNL, Universit\'e de Lyon, Universit\'e Lyon 1, CNRS/IN2P3, 69622 Villeurbanne cedex, France

 $^{3}$~CNRS-N\'eel, 38042 Grenoble Cedex 9, France

 $^{9}$~CSNSM, Univ. Paris-Sud, CNRS/IN2P3, Universit\'{e} Paris-Saclay, 91405 Orsay, France

 $^{5}$~Institute for Nuclear Research, MSP 03680 Kyiv, Ukraine

 $^{6}$~Karlsruhe Institute of Technology, Institut f\"{u}r Prozessdatenverarbeitung und Elektronik, 76021 Karlsruhe, Germany

 $^{7}$~Karlsruhe Institute of Technology, Institut f\"{u}r Experimentelle Kernphysik, 76128 Karlsruhe, Germany

 $^{8}$~Karlsruhe Institute of Technology, Institut f\"{u}r Kernphysik, 76021 Karlsruhe, Germany

 $^{9}$~Laboratory of Nuclear Problems, JINR, 141980 Dubna, Moscow region, Russia

 $^{10}$~IAS, CNRS, Universit\'e Paris-Sud, 91405 Orsay, France

 $^{11}$~University of Oxford, Department of Physics, Oxford OX1 3RH, U.K.

 $^{12}$~ICMCB, CNRS, Universit\'e de Bordeaux, 33608 Pessac Cedex, France

 $^{13}$~CEA, Centre d'Etudes Saclay, Orph\'ee, 91191 Gif-Sur-Yvette Cedex, France

 $^{14}$~Kirchhoff Institute for Physics, Heidelberg University, D-69120 Heidelberg, Germany

 $^{15}$~Dipartimento di Scienza e Alta Tecnologia dell'Universit\`a dell'Insubria, 22100 Como, Italy

 $^{16}$~INFN, Sezione di Milano-Bicocca, 20126 Milano, Italy

 $^{17}$~Nikolaev Institute of Inorganic Chemistry, 630090 Novosibirsk, Russia

 $^{18}$~Present address: Institut f\"{u}r Hochenergiephysik der \"{O}sterreichischen Akademie der Wissenschaften, A-1050 Wien, Austria

 $^{19}$~Department of Physics and Astronomy, University of Sheffield, Hounsfield Road, Sheffield S3 7RH, U.K.

 $^{20}$~CEA, LIST, LNHB, 91191 Gif-Sur-Yvette Cedex, France

 $^{21}$~CEA, Centre d'Etudes Saclay, IRAMIS, 91191 Gif-Sur-Yvette Cedex, France

 $^{22}$~Dipartimento di Fisica dell'Universit\'a di Milano-Bicocca, 20126 Milano, Italy

 $^{23}$~INFN, sezione di Roma, I-00185 Rome, Italy

 $^{24}$~CML Ltd., 630090 Novosibirsk, Russia

 }

\ead{danevich@kinr.kiev.ua}

\begin{abstract}

The LUMINEU is designed to investigate the possibility to search
for neutrinoless double beta decay in $^{100}$Mo by means of a
large array of scintillating bolometers based on ZnMoO$_4$
crystals enriched in $^{100}$Mo. High energy resolution and
relatively fast detectors, which are able to measure both the
light and the heat generated upon the interaction of a particle in
a crystal, are very promising for the recognition and rejection of
background events. We present the LUMINEU concepts and the
experimental results achieved aboveground and underground with
large-mass natural and enriched crystals. The measured energy
resolution, the $\alpha/\beta$ discrimination power and the
radioactive internal contamination are all within the
specifications for the projected final LUMINEU sensitivity.
Simulations and preliminary results confirm that the LUMINEU
technology can reach zero background in the region of interest
(around 3 MeV) with exposures of the order of hundreds
kg$\times$years, setting the bases for a next generation
$0\nu2\beta$ decay experiment capable to explore the inverted
hierarchy region of the neutrino mass pattern.

\end{abstract}

\section{Introduction}

Search for neutrinoless double beta decay ($0\nu2\beta$) is a
unique way to probe physics beyond the Standard Model since the
process violates the lepton number and requires the neutrino to be
a Majorana particle. The goal of the next generation $0\nu2\beta$
experiments is to test the inverted hierarchy of the neutrino mass
(the effective Majorana neutrino mass $\langle m_{\nu}\rangle$ is
at the level of a few hundredths of eV). The decay can be mediated
by many other effects beyond the Standard Model, like existence of
right-handed currents in weak interaction, Nambu-Goldstone bosons
(majorons), and many other hypothetical effects
\cite{Vergados:2012,Rodejohann:2012,Deppisch:2012,Bilenky:2015,Pas:2015}.

The isotope $^{100}$Mo is one of the most promising $2\beta$
nucleus taking into account the high energy of the decay
($Q_{2\beta}=3034.40(17)$ keV \cite{Rahaman:2008}), the
comparatively high natural isotopic abundance ($\delta=9.82(31)\%$
\cite{Berglund:2011}) and the possibility of isotopical separation
by centrifugation in an amount of hundreds kg - tons. The recent
calculations of nuclear matrix elements
\cite{Rodriguez:2010,Simkovic:2013,Hyvarinen:2015,Barea:2015} give
for $^{100}$Mo one of the shortest half-life in the range
$T^{0\nu2\beta}_{1/2}\approx(0.7-1.7)\times 10^{26}$ yr (for the
effective Majorana neutrino mass 0.05 eV, assume the value of the
axial vector coupling constant $g_{A}=1.27$ and using the
phase-space factor from \cite{Kotila:2012}).

Here we report the current status of the LUMINEU project
(Luminescent Underground Molybdenum Investigation for NEUtrino
mass and nature), aiming at preparing and fabricating of a next
generation $0\nu2\beta$ experiment to search for $0\nu2\beta$
decay of $^{100}$Mo at the level of the inverted hierarchy of
neutrino mass with cryogenic scintillating bolometers based on
zinc molybdate (ZnMoO$_4$) crystal scintillators enriched in
$^{100}$Mo \cite{Tenconi:2015}.
%The current status of the LUMINEU project is reported here.

\section{Development of ZnMoO$_4$ cryogenic scintillating bolometers}

The LUMINEU program requires a deep purification of molybdenum
both from radioactive elements and from impurities, which color
the ZnMoO$_4$ crystals \cite{Chernyak:2013}. A purification using
two-stage sublimation of molybdenum oxide in vacuum and
recrystallization from aqueous solutions of ammonium
para-molybdate was developed. A batch of LUMINEU crystals with
mass of the crystal boules up to 1.5 kg have been successfully
grown by the low-thermal-gradient Czochralski technique (LTG Cz),
and their optical, luminescent, diamagnetic, thermal and
bolometric properties were studied \cite{Berge:2014}. We have also
found that introducing tungsten oxide into the melt for crystal
growth at levels of fractions of percent improves the growth of
zinc molybdate crystals. The yield of the crystal boules, their
shape and overall quality becomes higher as a result of the melt
stabilization thanks to the prevention of the formation of an
extraneous phase in the synthesized ZnMoO$_4$ compound and in the
melt \cite{Chernyak:2015a}.

As a result of the R\&D the production cycle provides high quality
ZnMoO$_4$ scintillators with a high yield of the crystal boules
(more than 80\%) and low enough irrecoverable losses of molybdenum
(less than 4\%). These features are especially important for the
production of crystal scintillators from enriched materials. The
first zinc molybdate crystal with a mass of 0.17 kg was produced
from molybdenum enriched in $^{100}$Mo to 99.5\%
\cite{Barabash:2014}. Investigations of the scintillating
bolometers fabricated with two enriched crystals cut from the
boule show that the response of these devices meets the
requirements of a high-sensitivity double beta decay search based
on this technology, as discussed in \cite{Beeman:2012}. As a next
step, an enriched Zn$^{100}$MoO$_4$ crystal with a mass of 1.4 kg
was grown for the first time by the LTG Cz method
\cite{Poda:2015,Danevich:2015}.

\begin{figure}[h]
\includegraphics[width=17pc]{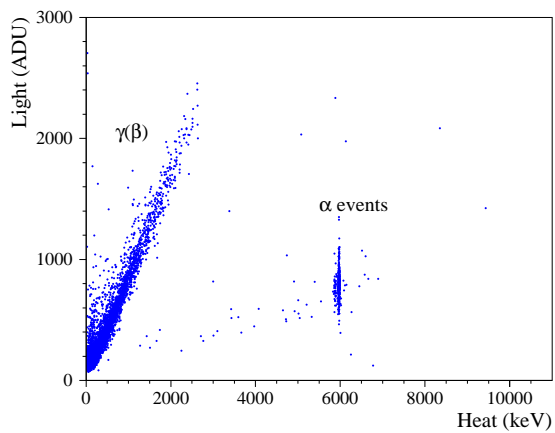}\hspace{2pc}
\begin{minipage}[b]{18pc}\caption{\label{scatter}Scatter plot of light versus heat signals accumulated with ZnMoO$_4$ scintillating bolometer with a mass of 334 g over 398 h in the EDELWEISS
set-up at the Modane underground laboratory. The $\beta$
($\gamma$) and $\alpha$ events are clearly separated. The energy
calibration used is drawn from the gamma quanta.}
\end{minipage}
\end{figure}

\begin{figure}[h]
\includegraphics[width=17pc]{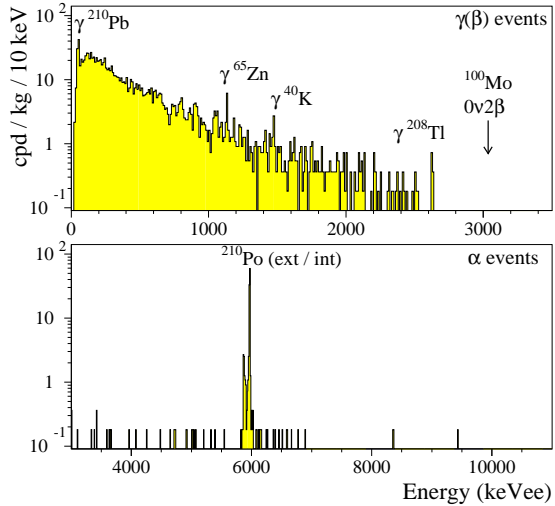}\hspace{2pc}%
\begin{minipage}[b]{18pc}\caption{\label{E-sp} The energy spectra of $\beta$ and $\gamma$
events (upper panel) and $\alpha$ events (lower panel) accumulated
with ZnMoO$_4$ scintillating bolometer over 398 h in the EDELWEISS
set-up at the Modane underground laboratory.}
\end{minipage}
\end{figure}

Several cryogenic scintillating bolometers were fabricated from
the produced crystal scintillators for aboveground and underground
tests. In the experimental set-ups the ZnMoO$_4$ crystal samples
were fixed inside copper holders by using PTFE supporting
elements. Each crystal was instrumented with temperature sensors
consisting of Neutron Transmutation Doped (NTD) Ge thermistors
attached at the crystals by using an epoxy glue. The light
detectors were developed from germanium disks and instrumented
with NTD sensors \cite{Tenconi:2012}. The light detectors were
mounted a few mm off the plane faces of the tested crystals
surrounded by reflective foil. The detectors with $\sim 0.3$ kg
ZnMoO$_4$ crystals have shown an excellent particle discrimination
ability (near 15$\sigma$) and a high energy resolution on the
level of 9 keV for the gamma quanta of $^{208}$Tl with energy 2615
keV (see Figs. \ref{scatter} and \ref{E-sp} where results of
underground measurements in the EDELWEISS set-up at the Modane
underground laboratory are presented).

Radioactive contamination of ZnMoO$_4$ crystal scintillators is
under test in the EDELWEISS set-up at the Modane underground
laboratory \cite{Armengaud:2015}. Preliminarily the ZnMoO$_4$
crystal scintillators show a very low radioactive contamination
(the results of one of the crystals test are presented in Table
\ref{radiopurity}) which is within the specifications for the
projected final LUMINEU sensitivity.

\begin{table}[h]
\caption{\label{radiopurity} Radioactive contamination of
ZnMoO$_4$ crystal scintillator with the mass of 334 g measured
over 2216 h in the EDELWEISS set-up at the Modane underground
laboratory \cite{Armengaud:2015}.}
\begin{center}
\begin{tabular}{ll}
\br
 Nuclide        & Activity (mBq/kg)    \\
 \mr
 $^{232}$Th     & $\leq 0.002$  \\
 $^{228}$Th     & $\leq 0.005$  \\
 $^{238}$U      & $\leq 0.002$  \\
% $^{230}$Th     & $\leq 0.002$  \\
 $^{226}$Ra     & $\leq 0.005$  \\
 $^{210}$Po     & $1.27(2)$     \\
 $^{235}$U      & $\leq 0.003$  \\
 $^{190}$Pt     & 0.004(1)      \\

\br
\end{tabular}
\end{center}
\end{table}

\section{Simulation of Zn$^{100}$MoO$_4$ detectors background}

\begin{figure}[h]
\includegraphics[width=17pc]{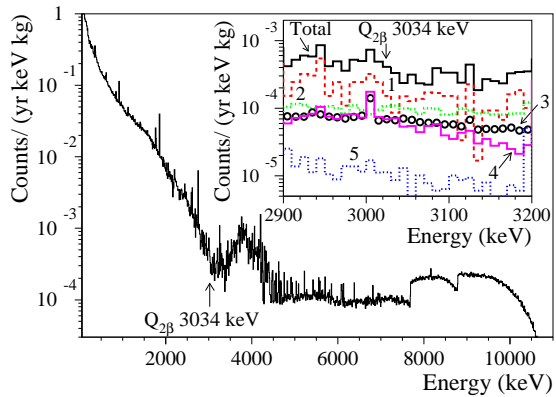}\hspace{2pc}%
\begin{minipage}[b]{18pc}\caption{\label{MC} The energy spectra of 48 Zn$^{100}$MoO$_4$ detectors in EDELWEISS set-up simulated with
the help of GEANT4 package. (Inset) The main components of background are shown: (1) surface contamination of Zn$^{100}$MoO$_4$ crystals,
(2) radioactive contamination of the reflecting foil surrounding the crystal, (3) bulk contamination of Zn$^{100}$MoO$_4$ crystals,
(4) contamination of copper holders and (5) PTFE clamps.}
\end{minipage}
\end{figure}

The background conditions of an experiment with Zn$^{100}$MoO$_4$
scintillating bolometers were estimated with the help of the
GEANT4 simulation package. We assume to locate 48
Zn$^{100}$MoO$_4$ detectors (with a mass $\sim 0.5$ kg each) in
the EDELWEISS set-up at the Modane underground laboratory. We
simulated contamination of Zn$^{100}$MoO$_4$ crystals and the
nearest materials to the detectors (copper holders, PTFE clamps,
reflective foil) by $^{238}$U and $^{232}$Th daughters. We have
also simulated cosmogenic nuclides in the Zn$^{100}$MoO$_4$
crystals and the copper holders. The simulated background energy
spectrum and its main components are shown in Fig. \ref{MC}. The
preliminary results of the simulations can be found in
\cite{Danevich:2015,Chernyak:2015b}.

The background due to random coincidence of events (especially of
$2\nu2\beta$ events \cite{Chernyak:2012}) can be reduced to the
level of $\approx 10^{-4}$ counts/(yr~keV~kg) with the help of
pulse-shape discrimination \cite{Chernyak:2014}. A total
background counting rate $\approx 5\times10^{-4}$
counts/(yr~keV~kg) can be reached, which corresponds to 3
counts/(yr~ton) in the region of interest (assuming 6 keV window
centered at the $0\nu2\beta$ peak position).

\section{Conclusions}

ZnMoO$_4$ based cryogenic scintillating bolometers, developed in
the framework of the LUMINEU project, show excellent performance:
a few keV energy resolution and $15\sigma$ alpha/beta particle
discrimination power at the $Q_{2\beta}$ value of $^{100}$Mo.
Radioactive contamination of ZnMoO$_4$ crystal scintillators
satisfies requirements for a large scale high sensitivity
experiment: activities of $^{228}$Th and $^{226}$Ra are less than
5 $\mu$Bq/kg. Enriched Zn$^{100}$MoO$_4$ crystals were grown for
the first time by using the LTG Cz technique. The production cycle
provided a high yield of the crystal boule ($\approx 80\%$ of the
initial charge) and an acceptable level of irrecoverable losses of
molybdenum ($\approx 4\%$). According to the Monte Carlo
simulation a background level of $\sim0.5$ counts/(yr~keV~ton) in
the region of interest can be reached in a large detector array.
These results pave the way to future sensitive searches based on
the LUMINEU technology, capable of exploring the inverted
hierarchy region of the neutrino mass pattern. The LUMINEU
activity is part of a CUPID, a proposed bolometric tonne-scale
experiment to be built as a follow-up of the CUORE experiment and
exploiting as much as possible the CUORE infrastructures
\cite{CUPID,CUPID-RD}.

\end{document}